\documentclass[11pt]{article}
\usepackage[margin=1in]{geometry}
\usepackage{amsmath, amssymb, amsthm}
\usepackage{algorithm}
\usepackage{algpseudocode}
\usepackage{booktabs}
\usepackage{graphicx}
\usepackage[authoryear]{natbib}
\usepackage{hyperref}
\usepackage[capitalize]{cleveref}

\newtheorem{theorem}{Theorem}
\newtheorem{lemma}[theorem]{Lemma}

\newcommand{\calM}{\ensuremath{\mathcal{M}}}
\newcommand{\calN}{\ensuremath{\mathcal{N}}}
\newcommand{\calV}{\ensuremath{\mathcal{V}}}
\newcommand{\bfC}{\ensuremath{\mathbf{C}}}
\newcommand{\bfZ}{\ensuremath{\mathbf{Z}}}
\newcommand{\bfc}{\ensuremath{\mathbf{c}}}
\newcommand{\bfv}{\ensuremath{\mathbf{v}}}
\newcommand{\bfx}{\ensuremath{\mathbf{x}}}
\newcommand{\bfy}{\ensuremath{\mathbf{y}}}
\newcommand{\bfz}{\ensuremath{\mathbf{z}}}
\newcommand{\ltwo}[1]{\left\|#1\right\|_2}
\newcommand{\boldzero}{\ensuremath{\boldsymbol{0}}}

\begin{document}

\title{Tractable Importance Sampling \\for Monte Carlo Privacy Accounting}
\author{
  Galen Andrew \\
  Google Research \\
  \texttt{galenandrew@google.com}
  \and
  Arun Ganesh \\
  Google Research \\
  \texttt{arunganesh@google.com}
}
\date{}

\maketitle

\begin{abstract}
Evaluating tight Differential Privacy (DP) parameters for DP matrix factorization with privacy amplification by sampling requires analyzing high-dimensional, correlated Gaussian mixtures. Because integrating these mixture distributions to find the exact privacy failure probability ($\delta$) is analytically intractable, these mechanisms are typically analyzed using \textit{Monte Carlo accounting}. However, standard sampling suffers from prohibitive variance when estimating small $\delta$ associated with strict privacy guarantees, which in turn requires a large sample complexity, making Monte Carlo accounting inefficient. We demonstrate that Monte Carlo accounting can be made much more efficient via \textit{importance sampling}. We give a shifted importance sampling proposal distribution parameterized by a stretch factor $\alpha > 1$. We prove a closed-form bound $B(\alpha) < 1$ on the support of the importance-weighted estimator that holds identically for both directions of the hockey-stick divergence, show that $B$ is log-convex so that the optimal $\alpha$ is a one-dimensional root-finding problem, and show that the resulting estimator seamlessly plugs into the KL-Hoeffding verification protocol used by past work, with a sample complexity reduced by exactly a factor of $1/B(\alpha^*)$.
\end{abstract}

\section{Introduction and Motivation}

A state-of-the-art technique for differentially private (DP) machine learning is \textit{DP matrix factorization} (DP-MF) \citep{denisov2022improved}, a generalization of DP stochastic gradient descent (DP-SGD) \citep{abadi2016deep} that adds noise correlated across each iteration of training (whereas DP-SGD adds independent noise in each iteration). Recent advances have combined these mechanisms with \textit{privacy amplification by sampling} \citep{kasiviswanathan2008what}, the observation that randomized batches increase uncertainty for the adversary, improving privacy.

For DP-SGD, the privacy can be analyzed for a single iteration, and then one can apply composition theorems to bootstrap this into a privacy guarantee for all iterations. The correlated noise in DP-MF rules out per-iteration composition.  \citet{choquettechoo2025nearexact} observed that \textit{Monte Carlo accounting}, which estimates the privacy parameter $\delta$ by drawing many samples from a distribution with mean $\delta$, can bypass the need for composition theorems. However, Monte Carlo accounting requires $\tilde{\Omega}(1/\delta)$ samples. $\delta$ is often set to be ``cryptographically'' small, i.e.\ less than $1 / |D|$ for dataset $D$, meaning Monte Carlo accounting has a very high sample complexity in practice.

In this work, we identify that Monte Carlo accounting of DP-MF is particularly amenable to \textit{importance sampling}, where we do Monte Carlo estimation using a shifted distribution rather than the distribution of interest, and reweight the samples so that the estimator remains unbiased for $\delta$. We propose an importance sampling method, and demonstrate that it can substantially reduce the sample complexity for Monte Carlo accounting.

In \cref{sec:background} we give background on DP-MF and Monte Carlo accounting and highlight why Monte Carlo accounting is particularly amenable to importance sampling. In \cref{sec:importancesampling} we propose and prove correctness of an importance sampling approach. In \cref{sec:samplecomplexity} we discuss the impact of our importance sampling approach on the sample complexity. In \cref{sec:implementation} we describe how to incorporate our importance sampling approach into a standard Monte Carlo verification pipeline, and some implementation details. Finally, in \cref{sec:experiments} we measure the sample complexity empirically.

\section{Background}\label{sec:background}

\subsection{DP-MF}

DP-MF is a generalization of DP-SGD parameterized by a strategy matrix $\bfC \in \mathbb{R}^{m \times m}$, where $m$ is the number of iterations in training. In each iteration of DP-SGD or DP-MF, we sample a batch, clip per-example gradients for each example in the batch, and then add Gaussian noise to their sum. In DP-SGD the Gaussian noise is independent in each iteration. In DP-MF, we first draw $\bfZ \in \mathbb{R}^{m \times d}$ with i.i.d.\ $\calN(0, \sigma^2)$ entries, where $d$ is the dimensionality of gradients, and add the $i$th row of $\bfC^{-1} \bfZ$ as the Gaussian noise. Note that $\bfC = \mathbb{I}$ recovers DP-SGD, i.e. all results we derive can also be applied to DP-SGD.

To analyze the privacy of DP-MF with randomized batches, fix an example in the dataset, and let $\bfx$ be the vector where $\bfx_i = 1$ iff that example participated in the batch for iteration $i$, and $0$ otherwise. Let $H_{e^\epsilon}(P, Q)$ be the $e^\epsilon$-hockey stick divergence, defined as:
\[H_{e^\epsilon}(P, Q) = \mathbb{E}_{y \sim P}[\max\{0, 1 - e^\epsilon Q(y) / P(y)\}].\]
\citet{choquettechoo2024privacyamplification} showed that using any randomized distribution of batches where each example participates independently from every other example, DP-MF satisfies the standard $(\epsilon, \delta)$-DP guarantee \citep{dwork2006calibrating} if $\max\{H_{e^\epsilon}(P, Q), H_{e^\epsilon}(Q, P)\} \leq \delta$, where
\[P = \sum_{\bfv} \Pr[\bfx = \bfv] \cdot \calN(\bfC \bfv, \sigma^2 \mathbb{I}), \qquad Q = \calN(\boldzero, \sigma^2 \mathbb{I}).\]

\subsection{Monte Carlo Accounting}

$H_{e^\epsilon}(P, Q)$ is difficult to evaluate because it is a high-dimensional integral over a mixture of Gaussians which in general is not easily decomposed. Monte Carlo accounting uses the observation that if we can sample $y \sim P$ and compute $Q(y)/P(y)$ efficiently, then we can do Monte Carlo estimation of $H_{e^\epsilon}(P, Q)$ by taking $y_1, \ldots, y_n \overset{\text{i.i.d.}}{\sim} P$ and computing
\[\hat{\delta} = \frac{1}{n} \sum_{i \in [n]} \max\left\{0, 1 - e^\epsilon \frac{Q(y_i)}{P(y_i)}\right\}.\]

Of course, this is only an estimate and by itself does not imply a formal privacy guarantee. \citet{wang2023randomized, dong2025bminsep} observed that we can get a high-probability tail bound on $\delta$ given $\hat{\delta}$, and absorb the failure probability of this tail bound into the privacy guarantee, getting an end-to-end formal privacy guarantee for Monte Carlo accounting plus running the downstream mechanism. See Appendix A of \citet{dong2025bminsep} for more details.

Unfortunately, estimating the mean of a distribution with mean $\approx \delta$, from individual samples potentially in $[0, 1]$ requires $\tilde{\Omega}(1/\delta)$ samples for reasonable confidence intervals, which is prohibitive for cryptographically small $\delta$.

\subsection{Importance Sampling}

Importance sampling provides variance reduction for Monte Carlo estimation. Suppose we want to estimate $\mathbb{E}_{y \sim P}[f(y)]$, as in Monte Carlo accounting. If a distribution $R$ is positive wherever $f(y) P(y) \neq 0$, then $\mathbb{E}_{y \sim R}[f(y) \cdot P(y)/R(y)] = \mathbb{E}_{y \sim P}[f(y)]$, but for carefully chosen $R$ the former may have significantly lower variance.

Monte Carlo accounting is an especially suitable candidate for importance sampling. To apply importance sampling to the hockey-stick divergence, we would need to define a distribution $R$ and then estimate:
\[H_{e^\epsilon}(P, Q) = \mathbb{E}_{y \sim R}\left[\max\left\{0, 1 - e^\epsilon \frac{Q(y)}{P(y)}\right\} \frac{P(y)}{R(y)}\right].\]

A key challenge in importance sampling is that the maximum of $P(y) / R(y)$ can be large, so an unwise choice of $R$ would actually increase the variance of the estimate. However, in the hockey-stick divergence the ``mask'' is of the form $\max\{0, 1 - e^\epsilon \cdot Q(y) / P(y)\}$. This term is $0$ if $P(y) / Q(y) \leq e^{\epsilon}$, so $P(y) / R(y)$ can be arbitrarily large in that region. We will demonstrate a choice of $R(y)$ that ensures the weighted estimand $\max\{0, 1 - e^\epsilon \cdot Q(y) / P(y)\} \cdot P(y) / R(y)$ is at most $B$ everywhere, for some $B < 1$. This reduces the range of our samples from $[0, 1]$ to $[0, B]$, immediately giving tighter confidence intervals and reduced sample complexity.

\section{Importance Sampling and Analysis}\label{sec:importancesampling}

In this section we propose and analyze a method to do importance sampling for Monte Carlo accounting of DP-MF. In this case, Monte Carlo accounting boils down to the following problem: Let $\bfc_i$ denote the $i$-th mode of the $T$-mode mixture of Gaussians $P$, as determined by DP-MF's strategy matrix $\bfC$ and the sampling strategy. For balls-in-bins sampling with cycle length $b$, the modes are indexed by the phase at which the example participates, so $T = b$ regardless of the number of iterations $m$; with a single epoch, $T = b = m$. We assume each mode is equally likely for notational simplicity; the argument goes through verbatim for non-uniform mode probabilities $\pi_i$, since the only step that uses the mixture structure is Jensen's inequality, which holds for any probability weights.\footnote{Replace every $\frac{1}{T}\sum_{i \in [T]}$ below by $\sum_{i \in [T]} \pi_i$.} We want to estimate
\begin{align}
    \delta_\text{fwd} &= \mathbb{E}_{\bfy \sim P}\left[\max\left\{0, 1 - e^{\epsilon} \frac{Q(\bfy)}{P(\bfy)}\right\}\right] \\
     \delta_\text{rev} &= \mathbb{E}_{\bfy \sim Q}\left[\max\left\{0, 1 - e^{\epsilon} \frac{P(\bfy)}{Q(\bfy)}\right\}\right],
\end{align}
where $P = \frac{1}{T} \sum_{i \in [T]} \calN(\bfc_i, \sigma^2 \mathbb{I})$ and $Q = \calN(\boldzero, \sigma^2 \mathbb{I})$. The analysis depends on the modes and the noise only through the largest mode signal-to-noise ratio
\begin{equation}\label{eq:kappa}
\kappa \triangleq \frac{\max_{i \in [T]} \ltwo{\bfc_i}^2}{\sigma^2}.
\end{equation}
For common use-cases such as ``column-normalized'' DP-MF with balls-in-bins sampling \citep{choquettechoo2025nearexact} or DP-SGD with random allocation \citep{dong2026optimal}, all modes have the same or very similar norms, so little is lost by taking the maximum.

To more commonly sample from the rare ``active region'' $P(\bfy) \ge e^\epsilon Q(\bfy)$ in the forward direction, we construct a proposal distribution $R_\text{fwd}$ by scaling the means of the mixture radially by a value $\alpha > 1$:
\begin{equation}
R_\text{fwd}(\bfy) = \frac{1}{T} \sum_{i \in [T]} \calN(\alpha \bfc_i, \sigma^2 \mathbb{I})
\end{equation}
We sample $\bfy_1, \dots, \bfy_n \sim R_\text{fwd}$ i.i.d.\ and construct the weighted estimator $\hat{\delta}_\text{fwd} = \frac{1}{n}\sum_{i=1}^n Y_i$, where
\begin{equation}
Y_i = \max\left\{0, 1 - e^\epsilon\frac{Q(\bfy_i)}{P(\bfy_i)}\right\} \frac{P(\bfy_i)}{R_\text{fwd}(\bfy_i)}.
\end{equation}
By the importance sampling identity, $\mathbb{E}_{\bfy_i \sim R_\text{fwd}}[Y_i] = \delta_\text{fwd}$. The reverse case is analogous.

\subsection{Unified Importance Sampling Bound}

To reveal the underlying geometric symmetry of both estimation problems, we introduce a general family of mixture distributions parameterized by a scalar $x$:
\[ P_x(\bfy) = \frac{1}{T} \sum_{i \in [T]} \calN(x \bfc_i, \sigma^2 \mathbb{I}) \]
Observe that our target distributions are exactly $P_1 = P$ and $P_0 = Q$. The forward and reverse accounting problems merely swap which of these two is the evaluation distribution ($P_v$) and which is the baseline ($P_u$). We parameterize our proposal distributions $R_\text{fwd}$ and $R_\text{rev}$ by geometrically extrapolating from the baseline through the evaluation distribution by a stretch factor $\alpha > 1$: $R_\text{fwd} = P_\alpha$, and $R_\text{rev} = P_{1 - \alpha}$.

Because $\{u, v\} = \{0, 1\}$, this geometric framing yields two algebraic invariants that hold for both directions. Let $s = v - u \in \{-1, 1\}$, and $w = u + \alpha s \in \{\alpha, 1-\alpha\}$.
\begin{enumerate}
    \item \textbf{Tail penalty invariance:} $w^2 - w = \alpha^2 - \alpha$, which is strictly positive since $\alpha > 1$.
    \item \textbf{Likelihood ratio invariance:} Expanding $v - sw$, we get \[ v - sw = v - su - \alpha s^2 = (v - uv + u^2) - \alpha = 1 - \alpha,\] unconditionally.
\end{enumerate}

\begin{lemma}[Unified Bound]\label{lem:unified}
    Let the evaluation distribution be $P_v$ and the baseline be $P_u$, where $\{u, v\} = \{0, 1\}$. For any proposal distribution $P_w$ where $w = u + \alpha(v - u)$ and $\alpha > 1$, the weighted estimand is globally bounded by:
    \[
    \max_\bfy \left[ \max\left\{0, 1  - e^\epsilon  \frac{P_u(\bfy)}{P_v(\bfy)}\right\} \frac{P_v(\bfy)}{P_w(\bfy)} \right] \leq B_{\kappa,\epsilon}(\alpha) \triangleq \frac{(\alpha - 1)^{\alpha - 1}}{\alpha^\alpha}\exp\left(\frac{\kappa(\alpha^2 - \alpha)}{2} - (\alpha - 1)\epsilon\right)
    \]

\end{lemma}
\begin{proof}
    We first upper bound the proposal density ratio subject to the likelihood ratio $P_v(\bfy) / P_u(\bfy) = e^c$. Expanding the PDF of $P_x(\bfy)$, we have:
    \[
    P_x(\bfy) \propto e^{-\ltwo{\bfy}^2/2\sigma^2} \sum_{i \in [T]} \exp\left(\frac{2x\langle \bfc_i, \bfy \rangle - x^2\ltwo{\bfc_i}^2}{2\sigma^2}\right)
    \]

    Let $r_i = \exp\left(\frac{2\langle \bfc_i, \bfy \rangle - \ltwo{\bfc_i}^2}{2\sigma^2}\right)$. For $x \in \{0, 1\}$, we have $x^2 = x$, so the sum simplifies to $\sum_{i \in [T]} r_i^x$. For the proposal $P_w$, we factor out the quadratic term in the exponent. Since $w^2 - w = \alpha^2 - \alpha > 0$ and $\ltwo{\bfc_i}^2 / \sigma^2 \leq \kappa$ by \eqref{eq:kappa}, we lower bound the denominator's terms to obtain:
    \[
        \frac{P_v(\bfy)}{P_w(\bfy)} = \frac{\sum_{i \in [T]} r_i^v}{\sum_{i \in [T]} r_i^w \exp\left(\frac{w - w^2}{2\sigma^2}\ltwo{\bfc_i}^2\right)} \leq \exp\left(\frac{\kappa(\alpha^2 - \alpha)}{2}\right) \frac{\frac{1}{T}\sum_{i \in [T]} r_i^v}{\frac{1}{T}\sum_{i \in [T]} r_i^w}
    \]

    Next, we apply the likelihood ratio condition $P_v(\bfy)/P_u(\bfy) = e^c$, which implies $\sum_i r_i^v / \sum_i r_i^u = e^c$. This reads $\frac{1}{T}\sum_i r_i = e^c$ when $(u, v) = (0, 1)$ and $\left(\frac{1}{T}\sum_i r_i\right)^{-1} = e^c$ when $(u, v) = (1, 0)$, so we can write $\frac{1}{T}\sum_i r_i = e^{sc}$. Now if $v=0$, we have $\frac{1}{T}\sum r_i^v = 1 = e^{c \cdot 0}$ while if $v=1$ we have $\frac{1}{T}\sum r_i^v = e^c = e^{c \cdot 1}$, so we can compactly write $\frac{1}{T}\sum_{i \in [T]} r_i^v = e^{cv}$.

    Because $\alpha > 1$, $w \notin [0,1]$, making the function $z \mapsto z^w$ strictly convex for $z > 0$. By Jensen's inequality:
    \[
    \frac{1}{T}\sum_{i \in [T]} r_i^w \geq \left(\frac{1}{T}\sum_{i \in [T]} r_i\right)^w = e^{scw}
    \]
    Substituting these into our ratio and applying the second invariant ($v - sw = 1 - \alpha$), we obtain:
    \[
    \frac{\frac{1}{T}\sum_{i \in [T]} r_i^v}{\frac{1}{T}\sum_{i \in [T]} r_i^w} \leq \frac{e^{cv}}{e^{scw}} = e^{c(v - sw)} = e^{-c(\alpha - 1)}
    \]
    Thus, conditional on the log-likelihood ratio $c$, the density ratio is bounded by $\exp\left(\frac{\kappa(\alpha^2 - \alpha)}{2} - c(\alpha - 1)\right)$.

    Finally, we maximize the full expression over all possible log-likelihoods $c \geq \epsilon$:
    \[
    \max_{c \geq \epsilon} \left[ (1 - e^{\epsilon - c})\exp\left(\frac{\kappa(\alpha^2 - \alpha)}{2} - c(\alpha - 1)\right)\right] = \exp\left(\frac{\kappa(\alpha^2 - \alpha)}{2}\right) \max_{c \geq \epsilon} \left[ e^{-c(\alpha - 1)} - e^{\epsilon - c\alpha} \right]
    \]
    Setting the derivative with respect to $c$ to zero yields the unique maximizer $c = \epsilon + \ln\left(\frac{\alpha}{\alpha - 1}\right)$. Plugging this optimal $c$ back into the expression yields the final global bound.
\end{proof}

The subscripts record that the bound depends on the problem only through $\kappa$ and $\epsilon$. It is increasing in $\kappa$ and, through the factor $e^{-(\alpha - 1)\epsilon}$, decreasing in $\epsilon$: $B_{\kappa,\epsilon'}(\alpha) \le B_{\kappa,\epsilon}(\alpha)$ for all $\epsilon' \ge \epsilon$. Since the importance weights $P_v(\bfy)/P_w(\bfy)$ do not depend on $\epsilon$, samples drawn with a stretch chosen for $\epsilon$ can therefore be reused to estimate $\delta$ at any larger $\epsilon'$ under the same support bound.

\subsection{Optimizing the Penalty and Importance Sampling}

With both the forward and reverse directions bounded by the exact same function $B_{\kappa,\epsilon}(\alpha)$, the remainder of our Monte Carlo accounting protocol is completely unified. For a given $\kappa$ and target privacy budget $\epsilon$, we numerically minimize $B_{\kappa,\epsilon}(\alpha)$ over $\alpha > 1$ to find the optimal geometric stretch factor $\alpha^*$ for our proposal distributions. Writing
\[ \log B_{\kappa,\epsilon}(\alpha) = (\alpha - 1)\ln(\alpha - 1) - \alpha \ln \alpha + \frac{\kappa(\alpha^2 - \alpha)}{2} - (\alpha - 1)\epsilon, \]
we have
\[ \frac{d}{d\alpha} \log B_{\kappa,\epsilon}(\alpha) = \ln\left(1 - \frac{1}{\alpha}\right) + \frac{\kappa(2\alpha - 1)}{2} - \epsilon,
\qquad
\frac{d^2}{d\alpha^2} \log B_{\kappa,\epsilon}(\alpha) = \frac{1}{\alpha(\alpha - 1)} + \kappa > 0. \]
Hence $B_{\kappa,\epsilon}$ is log-convex on $(1, \infty)$, and since $\frac{d}{d\alpha}\log B_{\kappa,\epsilon}(\alpha) \to -\infty$ as $\alpha \to 1^+$ while $\frac{d}{d\alpha}\log B_{\kappa,\epsilon}(\alpha) \to +\infty$ as $\alpha \to \infty$, there is a unique global minimizer $\alpha^* > 1$, which is the unique root of
\[ \ln\left(1 - \frac{1}{\alpha}\right) + \frac{\kappa(2\alpha - 1)}{2} = \epsilon. \]

Because $B_{\kappa,\epsilon}(\alpha) \to 1$ as $\alpha \to 1^+$, we recover the trivial bound $Y_i \leq 1$ of standard Monte Carlo (the proposal degenerates to the evaluation distribution), so $B_{\kappa,\epsilon}(\alpha^*) < 1$ strictly and importance sampling never increases the sample complexity. We use the proposal distribution $R_\text{fwd}=P_{\alpha^*}$ for the forward direction, and $R_\text{rev}=P_{1 - \alpha^*}$ for the reverse direction. Ignoring the $\ln(1 - 1 / \alpha)$ term in the derivative, we get a heuristic approximation $\alpha^* \approx \epsilon/\kappa + 1/2$. The $\ln (1 - 1/\alpha)$ term pushes it slightly past the threshold, further into the tail. Because of the log-convexity, we can easily use numerical optimization to nearly-exactly solve for $\alpha^*$ in practice.

\paragraph{When to use importance sampling.} Each importance-sampled draw costs about twice a standard one, since it requires the log-likelihood ratios of both $P$ and the proposal (\cref{sec:implementation}), so importance sampling is a net win in computation when $B_{\kappa,\epsilon}(\alpha^*) < 1/2$. Since $B_{\kappa,\epsilon}(\alpha^*)$ is a closed-form function of $(\kappa, \epsilon)$, this can be checked before drawing any samples. The bound approaches $1$ when $\alpha^* \to 1$, i.e.\ when $\kappa \gtrsim 2\epsilon$: a single mode is then already far from $(\epsilon, \delta)$-private on its own and essentially all of the privacy comes from amplification, which the bound does not see. In the experiments of \cref{sec:experiments}, $B_{\kappa,\epsilon}(\alpha^*)$ ranges from $0.56$ at the largest $\delta \approx 10^{-2}$ shown for $T = 1000$---the one point just past break-even---down to $10^{-6}$ at $\delta = 10^{-10}$.

For a single Gaussian ($T = 1$) the bound of Lemma~\ref{lem:unified} is attained with equality. Even for $P$ that are mixtures ($T > 1$), the use of Jensen's inequality is tight when all $r_i$ are equal; this can occur if e.g. the $\bfc_i$ are elementary vectors and $\bfy$ is a constant rescaling of the all-ones vector. So $B_{\kappa,\epsilon}(\alpha^*)$ cannot be improved without changing the proposal family or further assumptions on $P$, although it is typically far from tight where the proposal actually puts its mass (\cref{sec:variance}).

\section{KL-Hoeffding Confidence Intervals and Sample Complexity}\label{sec:samplecomplexity}

To turn our estimator $\hat{\delta} = \frac{1}{n} \sum_{i=1}^n Y_i$ into a rigorous guarantee, we apply concentration inequalities. Because $Y_i \in [0, B]$ with $B \triangleq B_{\kappa,\epsilon}(\alpha^*)$, we normalize the samples to lie in $[0,1]$ by defining $Z_i = Y_i / B$. Since $\mathbb{E}[Y_i] = \delta$, the scaled samples have mean $p = \delta / B$, and $\hat{p} = \hat{\delta} / B$ is their empirical mean.

For verification we need to control the probability of \emph{under}-estimating $\delta$: concluding that a mechanism is private when it is not. For i.i.d.\ $Z_i \in [0, 1]$ with mean $p$ and any $a < p$, the Chernoff--Hoeffding bound states
\begin{equation}
\mathbb{P}(\hat{p} \le a) \le \exp\left( -n D_{KL}(a \| p) \right), \qquad
D_{KL}(x \| y) = x \ln\left(\frac{x}{y}\right) + (1-x)\ln\left(\frac{1-x}{1-y}\right).
\end{equation}
This is the same inequality that underlies the Monte Carlo verification protocol of \citet{dong2025bminsep}, which the estimator plugs into unchanged: we test whether $\hat{\delta} \le \delta_\text{base}$ for a verification threshold $\delta_\text{base}$, and for any $\tau \in [1, B / \delta_\text{base}]$, a mechanism with true $\delta > \tau \delta_\text{base}$ passes the test with probability at most
\begin{equation}
q(\tau) = \exp\left(-n \, D_{KL}\!\left(\frac{\delta_\text{base}}{B} \,\middle\|\, \frac{\tau \delta_\text{base}}{B}\right)\right),
\end{equation}
so the end-to-end pipeline satisfies $(\epsilon, \tau \delta_\text{base} + q(\tau)(B - \tau\delta_\text{base}))$-DP, optimized over $\tau$. Here $B$ replaces the trivial bound of $1$ used for standard Monte Carlo in the failure branch, since $\delta = \mathbb{E}[Y_i] \le B$ holds regardless of the outcome of the verification; this is not a negligible refinement, as $q(\tau)$ is of the same order as $\delta_\text{base}$ at the optimum. Factoring out $B$, the reported delta is $B \cdot [\tau (\delta_\text{base}/B) + q(\tau)(1 - \tau \delta_\text{base}/B)]$. In other words, the protocol for importance-sampled estimates is exactly the protocol for standard Monte Carlo estimates with every $\delta$ replaced by $\delta / B$. (If $B$ exceeds our target privacy parameter $\delta$, then the upper bound $B$ itself certifies the privacy guarantee and no Monte Carlo estimation is needed.)

The KL form of the bound is essential. A plain Hoeffding bound $\exp(-2 n t^2)$ with $t = (\tau - 1)\delta_\text{base}/B$ would require $n = \Theta(B^2 / \delta_\text{base}^2)$ samples, which is worse than standard Monte Carlo. The KL form, by contrast, exploits the fact that a $[0,1]$-valued variable with a tiny mean also has tiny variance. For $y \ll 1$ one has $D_{KL}(x \| y) = x \ln(x / y) - x + y + O(y^2)$, so with $x = \delta_\text{base}/B$ and $y = \tau\delta_\text{base}/B$,
\begin{equation}
n \, D_{KL}\!\left(\frac{\delta_\text{base}}{B} \,\middle\|\, \frac{\tau \delta_\text{base}}{B}\right) \approx \frac{n}{B} \cdot \delta_\text{base} \left(\tau - 1 - \ln \tau\right).
\end{equation}
The right-hand side is precisely the exponent that standard Monte Carlo (where $B = 1$) would achieve with $n / B$ samples. Thus $n$ importance-sampled draws with support bound $B$ are worth $n / B$ standard draws: the sample complexity improves by a factor of exactly $1 / B_{\kappa,\epsilon}(\alpha^*)$, which can be many orders of magnitude.\footnote{The second-order Taylor expansion $D_{KL}(x \| y) \approx (x - y)^2 / 2y$ is only valid for $|x - y| \ll y$, i.e.\ $\tau - 1 \ll 1$, which is not the regime of interest ($\tau$ is a constant like $2$). In practice one should simply evaluate the exact binary KL divergence; the approximation above is only to build intuition.}

\paragraph{Choosing $\tau$.} Since importance sampling makes samples cheap, one might hope to spend more of them on a smaller slack $\tau$, so that the reported $\delta_\text{total} \approx \tau \delta_\text{base}$ is closer to $\delta_\text{base}$ and less noise is needed. The trade is poor. The sample count scales as $1 / (\tau - 1 - \ln \tau) \approx 2 / (\tau - 1)^2$, so relative to $\tau = 2$, reaching $\tau = 1.2$ costs $17\times$ the samples and $\tau = 1.05$ costs $250\times$; but $\tau \to 1$ can at most halve $\delta_\text{total}$, and the noise multiplier depends on $\delta$ only logarithmically, so even $\tau = 1.05$ reduces $\sigma$ by only $1$--$3\%$ for the Gaussian mechanism at $\delta \in [10^{-10}, 10^{-6}]$. We use $\tau = 2$ throughout; $\tau \in [1.5, 2]$ is a sensible operating range.

\section{Monte Carlo Evaluation Algorithm}\label{sec:implementation}

Because the mixture distribution $P$ and the baseline distribution $Q$ are asymmetric in shape, their expected hockey-stick divergences are not identical. To rigorously establish $(\epsilon, \delta)$-DP, we must compute both the forward and reverse privacy losses and take the maximum. We utilize the optimal geometric stretch factor $\alpha^*$ to construct the forward proposal $P_{\alpha^*}$ and the symmetric reverse proposal $P_{1-\alpha^*}$.

\begin{algorithm}[H]
\caption{Importance Sampling for Monte Carlo Accounting of DP-MF}
\label{alg:is}
\begin{algorithmic}[1]
\Require Samples $n$, Target $\epsilon$, Noise scale $\sigma$, Strategy Matrix modes $\{\bfc_1, \dots, \bfc_T\}$, Boolean \textbf{reverse}
\State $\kappa \gets \max_k \ltwo{\bfc_k}^2 / \sigma^2$ \Comment{Eq.~\eqref{eq:kappa}}
\State Numerically minimize unified bound $B_{\kappa,\epsilon}(\alpha)$ over $\alpha > 1$ to find optimal stretch $\alpha^*$
\State Calculate strict upper bound $B \gets B_{\kappa,\epsilon}(\alpha^*)$

\If{\textbf{reverse}}
    \State $\beta \gets 1 - \alpha^*$
    \State $P(\bfy) \gets \mathcal{N}(\bfy; \boldzero, \sigma^2 \mathbb{I})$
    \State $Q(\bfy) \gets \frac{1}{T} \sum_{k=1}^T \mathcal{N}(\bfy; \bfc_k, \sigma^2 \mathbb{I})$
\Else
    \State $\beta \gets \alpha^*$
    \State $P(\bfy) \gets \frac{1}{T} \sum_{k=1}^T \mathcal{N}(\bfy; \bfc_k, \sigma^2 \mathbb{I})$
    \State $Q(\bfy) \gets \mathcal{N}(\bfy; \boldzero, \sigma^2 \mathbb{I})$
\EndIf
\State $R(\bfy) \gets \frac{1}{T} \sum_{k=1}^T \mathcal{N}(\bfy; \beta \bfc_k, \sigma^2 \mathbb{I})$

\Statex
\State Initialize $Y \gets 0$
\For{$i = 1 \dots n$}
    \State Draw uniform discrete component $j \sim \text{Uniform}(1, T)$
    \State Draw continuous noise $\bfz \sim \mathcal{N}(\boldzero, \mathbb{I})$
    \State $\bfy \gets \beta \bfc_j + \sigma \bfz$
    \State $Y \gets Y + \max\left\{0, 1 - e^\epsilon \frac{Q(\bfy)}{P(\bfy)}\right\} \times \frac{P(\bfy)}{R(\bfy)}$
\EndFor

\State $\hat{\delta} \gets Y / n$
\State \Return Empirical $\hat{\delta}$, and scaling bound $B$.
\end{algorithmic}
\end{algorithm}

\paragraph{Implementation remarks.} All three densities $P$, $Q$, $R$ are of the form $P_x$ and share the factor $\calN(\bfy; \boldzero, \sigma^2\mathbb{I})$, so the per-sample computation only requires the log-likelihood ratios $\ell_x(\bfy) \triangleq \log P_x(\bfy) / P_0(\bfy)$ for $x \in \{1, \beta\}$; these are exactly the privacy-loss computations already used for standard Monte Carlo accounting, applied with the modes scaled by $x$. In the forward direction the summand is $\exp\big(\log(1 - e^{\epsilon - \ell_1(\bfy)}) + \ell_1(\bfy) - \ell_\beta(\bfy)\big)$ when $\ell_1(\bfy) > \epsilon$ (and $0$ otherwise); in the reverse direction it is $\exp\big(\log(1 - e^{\epsilon + \ell_1(\bfy)}) - \ell_\beta(\bfy)\big)$ when $\ell_1(\bfy) < -\epsilon$. Finally, Lemma~\ref{lem:unified} guarantees every summand lies in $[0, B]$, which serves as a free runtime consistency check on the implementation. For balls-in-bins sampling the modes are shifts of one another ($\bfc_j$ is $\bfc_0$ shifted down by $j$ rounds and truncated), so the inner products $\langle \bfc_j, \bfy \rangle$ for all $T$ modes are a single cross-correlation of $\bfy$ with $\bfc_0$, computable by FFT in $O((m + T) \log(m + T))$ time per sample rather than the $O(mT)$ of multiplying by the dense mode matrix, which is mostly zeros for banded $\bfC$.

\subsection{Calibrating with Importance Sampling}

\begin{algorithm}[t]
\caption{Calibration via Monte Carlo Accounting}
\label{alg:calibration}
\textbf{Parameters:} Candidate mechanisms $\calM_1, \calM_2, \calM_3, \ldots, \calM_k$, verifier $\calV$ with outputs $0, 1$.
\begin{algorithmic}[1]
\Require $\calM_1$ is known to satisfy the target privacy guarantee without verification.
\For{$i \in \{2, \ldots, k\}$}
\State $v_i \leftarrow \calV(\calM_{i})$
\If{$v_i = 0$}
\State Run $\calM_{i-1}$ and \Return its output.
\EndIf
\EndFor
\State Run $\calM_k$ and \Return its output.
\end{algorithmic}
\end{algorithm}

\citet{dong2025bminsep} gave a method for converting primitives for Monte Carlo estimation into a procedure for calibrating a parameter (such as the noise parameter) of a target DP mechanism, such that the end-to-end process of calibration + running the target mechanism satisfies $(\epsilon, \delta)$-DP. The idea is to define a discrete sweep over the parameter. We iterate over the parameters from most to least private (e.g., largest to smaller noise). For each parameter, we use Monte Carlo estimation as a black-box verifier for whether that parameter achieves the target privacy guarantee. As soon as the verifier reports that the target parameter doesn't achieve the target privacy guarantee, we report the previous value in the sweep should be used for the target mechanism. We restate their algorithm formally in \cref{alg:calibration}. The first candidate $\calM_1$ is never verified: it is assumed to satisfy the target guarantee without Monte Carlo accounting, e.g.\ because its noise is calibrated analytically, and it is the fallback when even $\calM_2$ fails verification. (Our implementation verifies every candidate and returns a failure flag if the first one fails, leaving the choice of fallback to the caller; this is the same algorithm with the fallback playing the role of $\calM_1$.)

Their analysis effectively assumes that if the verification passes when it should fail, then w.p. 1 a non-private output is published. However, if we have a support bound $B$ on each mechanism's Monte Carlo samples, then we know the mechanisms must satisfy $(\epsilon, B)$-DP, letting us slightly tighten their analysis:

\begin{lemma}\label{lem:calibration}
    Suppose for each mechanism $\calM_i$ that if the mechanism does not satisfy $(\epsilon, \delta)$-DP, then $\calV(\calM_i) = 0$ w.p. at least $1 - q$. Furthermore, assume for some $B \geq \delta$ that each $\calM_i$ is $(\epsilon, B)$-DP, and additionally that $\calM_1$ is $(\epsilon, \delta)$-DP. Then \cref{alg:calibration} is $(\epsilon, \delta + (B- \delta)q)$-DP.
\end{lemma}
With the verifier of \cref{sec:samplecomplexity}, $\delta = \tau \delta_\text{base}$ and $q = q(\tau)$, and the lemma recovers the guarantee $(\epsilon, \tau \delta_\text{base} + q(\tau)(B - \tau\delta_\text{base}))$-DP stated there.
\begin{proof}
    We assume at least one of the mechanisms $\calM_i$ does not satisfy $(\epsilon, \delta)$-DP, otherwise \cref{alg:calibration} is a convex combination of $(\epsilon, \delta)$-DP mechanisms and thus satisfies $(\epsilon, \delta)$-DP.

    Let $\calM_{i^*}$ be the (randomly chosen) mechanism that is run by \cref{alg:calibration}.  Let $\ell$ be the first ``bad'' index, i.e. the index such that $\calM_{\ell}$ does not satisfy $(\epsilon, \delta)$-DP, but for any $i < \ell$ $\calM_i$ does satisfy $(\epsilon, \delta)$-DP (since we assume $\calM_1$ satisfies $(\epsilon, \delta)$-DP and some other $\calM_i$ does not, $\ell$ must exist). Note that by the assumption on $\calV$, we have $\Pr[i^* \geq \ell] \leq q$. Let $\calM$ be the overall mechanism given by \cref{alg:calibration}. Since the randomness of $\calV$ is independent of the data, $i^*$ has the same distribution under $D$ and $D'$; in particular, conditioned on any event defined by $i^*$, $\calM$ is a convex combination of the $\calM_i$ with data-independent weights. Then we have for any $D \simeq D'$ and output set $S$:

\begin{align*}
    \Pr_{y \sim \calM(D)}[y \in S] =& \Pr_{y \sim \calM(D)}[y \in S | i^* < \ell] \cdot \Pr[i^* < \ell] + \Pr_{y \sim \calM(D)}[y \in S | i^* \geq \ell] \cdot \Pr[i^* \geq \ell]\\
    \leq& \left(e^\epsilon \Pr_{y \sim \calM(D')}[y \in S | i^* < \ell] + \delta\right) \cdot \Pr[i^* < \ell] \\
    &+ \left(e^\epsilon\Pr_{y \sim \calM(D')}[y \in S | i^* \geq \ell] + B \right)\cdot \Pr[i^* \geq \ell]\\
    =&e^\epsilon \Pr_{y \sim \calM(D')}[y \in S] + \delta \cdot \Pr[i^* < \ell] + B \cdot \Pr[i^* \geq \ell]\\
    \leq&e^\epsilon \Pr_{y \sim \calM(D')}[y \in S] + \delta + (B - \delta)q.
\end{align*}

The first inequality uses the fact that conditioned on $i^* < \ell$, $\calM$ is a convex combination of $(\epsilon, \delta)$-DP mechanisms, and conditioned on $i^* \geq \ell$ is a convex combination of $(\epsilon, B)$-DP mechanisms. The second uses the bound on $\Pr[i^* \geq \ell]$ and the fact $B \geq \delta$.

\end{proof}

It is important to note that each individual mechanism $\calM_i$ might have a different support bound $B$ when using importance sampling. The variable support bounds interact with calibration in two different ways in \cref{lem:calibration}. The additional error term $(B - \delta)q$ uses the \textit{maximum} across all support bounds. This maximum is necessary: in the worst case, only $\calM_k$ does not satisfy $(\epsilon, \delta)$-DP, and also has the worst support bound, in which case the analysis of \cref{lem:calibration} could be tight. Refining this bound by validating that we are not in the worst case would require knowledge of which mechanisms satisfy the target privacy guarantee, which is the entire point of running \cref{alg:calibration} in the first place.

However, the verifier is a black box, and we only care about its failure probability. This means if we use Monte Carlo estimation as the verifier, we can compute the failure probability for each verifier using its support bound (and in turn, possibly using a different number of Monte Carlo samples for each verifier), even if our final privacy guarantee uses the maximum of the support bounds.

\section{Bounding the Reverse Direction Analytically}\label{sec:reverse}

The reverse divergence $\delta_\text{rev} = H_{e^\epsilon}(Q, P)$ is active only where the \emph{whole} mixture is unlikely relative to $Q$, which for well-separated modes is far rarer than the forward active region. The following lemma makes this precise and gives a closed-form upper bound on $\delta_\text{rev}$ that requires no sampling. Let $\bar{\bfc} = \frac{1}{T}\sum_{i \in [T]} \bfc_i$ be the mean mode, $\mu = \frac{\ltwo{\bar{\bfc}}}{\sigma}$, and
\[ \Delta \triangleq \frac{\frac{1}{T}\sum_{i \in [T]} \ltwo{\bfc_i}^2 - \ltwo{\bar{\bfc}}^2}{2\sigma^2} \;\ge\; 0. \]

\begin{lemma}[Reverse direction]\label{lem:reverse}
For every $\epsilon$,
\[ H_{e^\epsilon}(Q, P) \;\le\; H_{e^{\epsilon - \Delta}}\big(Q, \calN(\bar{\bfc}, \sigma^2 \mathbb{I})\big)
= \Phi\!\left(\frac{\mu}{2} - \frac{\epsilon - \Delta}{\mu}\right) - e^{\epsilon - \Delta}\,\Phi\!\left(-\frac{\mu}{2} - \frac{\epsilon - \Delta}{\mu}\right), \]
where $\Phi$ is the standard normal CDF.
\end{lemma}
\begin{proof}
With $r_i$ as in the proof of \cref{lem:unified}, $P(\bfy)/Q(\bfy) = \frac{1}{T}\sum_i r_i \ge \big(\prod_i r_i\big)^{1/T}$ by the AM--GM inequality, and
\[ \Big(\prod_i r_i\Big)^{1/T} = \exp\left(\frac{2\langle \bar{\bfc}, \bfy\rangle - \frac{1}{T}\sum_i \ltwo{\bfc_i}^2}{2\sigma^2}\right) = e^{-\Delta}\,\frac{\calN(\bar{\bfc}, \sigma^2\mathbb{I})(\bfy)}{Q(\bfy)}. \]
The integrand $x \mapsto \max\{0, 1 - e^\epsilon x\}$ of $H_{e^\epsilon}(Q, P) = \mathbb{E}_{\bfy \sim Q}[\max\{0, 1 - e^\epsilon P(\bfy)/Q(\bfy)\}]$ is nonincreasing, so replacing $P/Q$ by this lower bound can only increase the expectation, giving the first inequality; the equality is the standard formula for the hockey-stick divergence between two Gaussians with the same covariance. $\Delta \ge 0$ by Jensen's inequality.
\end{proof}

For $T = 1$ the lemma holds with equality (the Gaussian pair is symmetric), while for many modes pointing in different directions $\ltwo{\bar{\bfc}}$ is small---for balls-in-bins over one epoch with a nonnegative $b$-banded $\bfC$ with unit-norm columns, $\ltwo{\bar{\bfc}}^2 \le b/T$---so the bound is tiny. The forward direction admits no such bound, since its integrand is increasing in $P/Q$; this asymmetry is why $\delta_\text{fwd}$ must be estimated by sampling. The bound is informative only when $\epsilon > \Delta$, i.e.\ when a single mode's signal-to-noise ratio $\kappa$ is below roughly $2\epsilon$, which is also the condition under which importance sampling is worthwhile (\cref{sec:importancesampling}). In that regime it lies far below $\delta_\text{fwd}$---in all experiments of \cref{sec:experiments} by a factor of at least $29$---and wherever it is below the verification threshold $\delta_\text{base}$ of \cref{sec:samplecomplexity}, \cref{alg:is} need only be run in the forward direction, halving the number of samples. At small $\epsilon$ with few modes the two directions are genuinely comparable (for $\epsilon = 1$ and $T = 10$ we measured $\delta_\text{rev} \approx 0.2$--$0.6\,\delta_\text{fwd}$), and both must be sampled.

\section{Empirical Sample Complexity}\label{sec:experiments}

We now measure how many samples the protocol of \cref{sec:samplecomplexity} needs in practice. For a mixture, $\delta$ has no closed form, so each point is located empirically: we run \cref{alg:is} in both directions until the estimate $\hat{\delta}$ has a relative standard error below $10\%$, and then report the number of samples $n$ the KL-Hoeffding protocol needs to certify $\delta = \hat{\delta}$ with $\tau = 2$ and failure probability $q = \delta$, so that the reported $\delta_\text{total} = \tau\delta + q(B - \tau\delta)$ is at most $(2 + B)\delta \le 3\delta$; explicitly, $n = \ln(1/\delta) / D_{KL}(\delta / B \,\|\, 2\delta / B)$. For standard Monte Carlo the same formula applies with $B = 1$. We call this the number of \emph{samples to certify} $\delta$, as opposed to the number of samples needed merely to estimate it to a given precision.

\begin{figure}[t]
\centering
\includegraphics[width=0.75\textwidth]{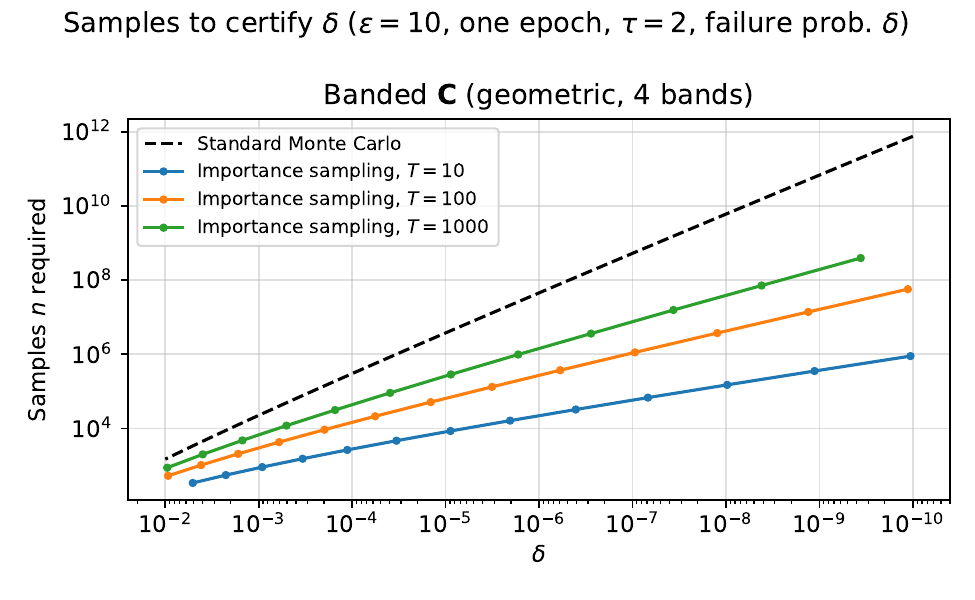}
\caption{Samples needed to certify $(\epsilon, \delta)$-DP with $\epsilon = 10$ for balls-in-bins sampling with $T \in \{10, 100, 1000\}$ modes over a single epoch, as a function of $\delta$ (swept via the noise multiplier), for a banded $\bfC$ whose column is the normalized geometric sequence $(1, \frac12, \frac14, \frac18)$. Dashed: standard Monte Carlo ($B = 1$). Solid: importance sampling with the support bound $B_{\kappa,\epsilon}(\alpha^*)$. Each point is estimated to $10\%$ relative standard error.}
\label{fig:mixtures}
\end{figure}

\Cref{fig:mixtures} shows the results for $\epsilon = 10$ and one epoch ($m = T$). We make two observations.

\paragraph{Gains are large and grow as $\delta$ shrinks.} At the smallest $\delta$ shown ($\approx 10^{-10}$), importance sampling reduces the number of samples to certify by factors of roughly $8 \cdot 10^5$ ($T = 10$), $10^4$ ($T = 100$), and $800$ ($T = 1000$). The gain is exactly $1 / B_{\kappa,\epsilon}(\alpha^*)$ (\cref{sec:samplecomplexity}), and $B$ decreases with the noise multiplier, so the gain grows as $\delta$ shrinks. In every run the reverse direction $\delta_\text{rev}$ was negligible---zero on all samples except at the largest $\delta$ for $T = 10$, where it was $10^{-4}$ of $\delta_\text{fwd}$---so $\delta_\text{fwd}$ determined every point.

\paragraph{The gain is smaller for larger $T$.} The support bound $B$ depends on the mixture only through $\kappa$: it knows how heavy the tail of a \emph{single} Gaussian mode is, but nothing about privacy amplification, which is what makes the mixture's $\delta$ much smaller than a single mode's. Importance sampling therefore removes the cost of the Gaussian tail---$B$ is within a small constant of a single mode's hockey-stick divergence, so that part of the sample count becomes $O(1)$---but the amplification factor, the ratio between a single mode's $\delta$ and the mixture's, is paid in full by standard Monte Carlo and importance sampling alike. Since more modes means more amplification, the remaining cost, and hence the number of samples to certify at a given $\delta$, grows with $T$. (Heuristically, near a mode the mixture likelihood ratio is about $T$ times smaller than the single-mode ratio, so the mixture at privacy budget $\epsilon$ behaves like a single mode at $\epsilon + \ln T$; the amplification factor then grows polynomially in $T$, with an exponent between $\alpha^* - 1$ and $\alpha^*$.)

\subsection{Looseness of the support bound}\label{sec:variance}

The support bound $B$ enters the protocol as a proxy for the variance of the weighted samples: a $[0, B]$-valued variable with mean $\delta$ has variance at most $B \delta$, and the KL-Hoeffding bound is tight for the variable that attains it. To see how far the actual estimand is from this worst case, \cref{tab:variance} reports, for three points on each curve of \cref{fig:mixtures}, the empirical variance $\hat{v}$ of the weighted samples relative to $B \hat{\delta}$, together with the number of samples $100 \hat{v} / \hat{\delta}^2$ needed to \emph{estimate} $\hat{\delta}$ to $10\%$ relative standard error and the number of samples to \emph{certify} it.

\begin{table}[t]
\centering
\begin{tabular}{rrrrrrr}
\toprule
$T$ & $\sigma$ & $\hat{\delta}$ & $B$ & $B \hat{\delta} / \hat{v}$ & $n_\text{est}$ & $n_\text{cert}$\\
\midrule
$10$ & $0.32$ & $5.1 \cdot 10^{-3}$ & $1.1 \cdot 10^{-1}$ & $6.8$ & $3.1 \cdot 10^{2}$ & $3.4 \cdot 10^{2}$\\
$10$ & $0.44$ & $2.0 \cdot 10^{-6}$ & $7.8 \cdot 10^{-4}$ & $56$ & $6.7 \cdot 10^{2}$ & $1.6 \cdot 10^{4}$\\
$10$ & $0.57$ & $1.1 \cdot 10^{-10}$ & $1.3 \cdot 10^{-6}$ & $918$ & $1.3 \cdot 10^{3}$ & $8.9 \cdot 10^{5}$\\
\midrule
$100$ & $0.27$ & $9.3 \cdot 10^{-3}$ & $3.4 \cdot 10^{-1}$ & $7.1$ & $5.1 \cdot 10^{2}$ & $5.3 \cdot 10^{2}$\\
$100$ & $0.40$ & $5.9 \cdot 10^{-7}$ & $4.7 \cdot 10^{-3}$ & $356$ & $2.2 \cdot 10^{3}$ & $3.7 \cdot 10^{5}$\\
$100$ & $0.49$ & $1.1 \cdot 10^{-10}$ & $8.6 \cdot 10^{-5}$ & $12{,}209$ & $6.2 \cdot 10^{3}$ & $5.7 \cdot 10^{7}$\\
\midrule
$1000$ & $0.25$ & $9.5 \cdot 10^{-3}$ & $5.6 \cdot 10^{-1}$ & $5.8$ & $1.0 \cdot 10^{3}$ & $8.7 \cdot 10^{2}$\\
$1000$ & $0.35$ & $1.7 \cdot 10^{-6}$ & $3.8 \cdot 10^{-2}$ & $327$ & $6.9 \cdot 10^{3}$ & $9.8 \cdot 10^{5}$\\
$1000$ & $0.42$ & $3.6 \cdot 10^{-10}$ & $2.0 \cdot 10^{-3}$ & $17{,}245$ & $3.2 \cdot 10^{4}$ & $3.9 \cdot 10^{8}$\\
\bottomrule
\end{tabular}
\caption{Empirical variance of the importance-weighted samples for the configurations of \cref{fig:mixtures}, at the largest, an intermediate, and the smallest $\delta$ of each curve. $B \hat{\delta} / \hat{v}$ is the factor by which the worst-case variance implied by the support bound exceeds the observed variance; $n_\text{est} = 100\hat{v}/\hat{\delta}^2$ is the number of samples needed to estimate $\hat{\delta}$ to $10\%$ relative standard error, and $n_\text{cert}$ the number of samples to certify it ($\tau = 2$, failure probability $\delta$).}
\label{tab:variance}
\end{table}

\paragraph{The estimator is far more precise than the certificate.} At large $\delta$, $n_\text{est}$ and $n_\text{cert}$ nearly coincide for every $T$ (e.g.\ $3.1 \cdot 10^2$ versus $3.4 \cdot 10^2$ for $T = 10$), but the ratio $B \hat{\delta} / \hat{v}$ grows steadily as $\delta$ shrinks, reaching $10^3$--$2 \cdot 10^4$ at $\delta \approx 10^{-10}$, where estimating $\hat{\delta}$ to $10\%$ takes $10^3$--$3 \cdot 10^4$ samples while certifying it takes $10^6$--$4 \cdot 10^8$. The Jensen step of \cref{lem:unified} is tight only when all $r_i$ are equal, and is far from tight where the proposal actually puts its mass. This headroom cannot, however, be recovered by any verification protocol that relies on the support bound alone, including one that uses the empirical variance. Consider the estimand that takes the value $B$ with probability $\tau \delta_\text{base} / B$ and $0$ otherwise: it has mean $\tau \delta_\text{base}$ and so must be rejected, yet with $n \ll B / (\tau \delta_\text{base})$ samples it produces all zeros, with empirical mean and variance both $0$, exactly as a far more private mechanism would. Any certificate valid for all $[0, B]$-valued estimands therefore needs $n \gtrsim B \ln(1/q) / (\tau \delta_\text{base})$ samples, within a small constant of the KL-Hoeffding count: given only the support bound, the protocol of \cref{sec:samplecomplexity} is essentially optimal.

\subsection{Optimized banded matrix}

\begin{figure}[t]
\centering
\includegraphics[width=0.6\textwidth]{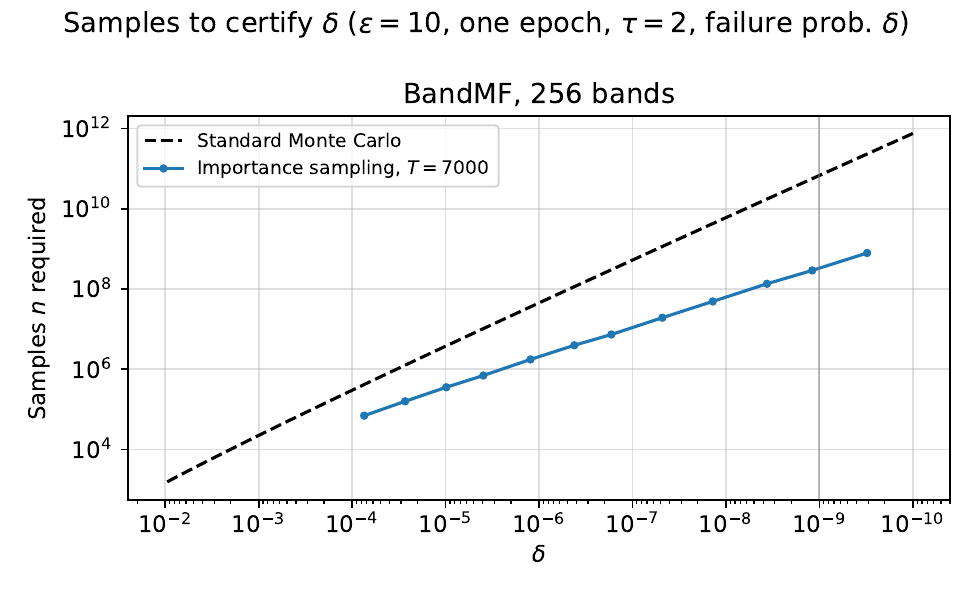}
\caption{Samples needed to certify $\delta$ for the optimized banded configuration ($m = T = 7000$, $256$ bands, $\epsilon = 10$). Curves as in \cref{fig:mixtures}; the vertical line marks the target $\delta = 10^{-9}$.}
\label{fig:banded256}
\end{figure}

We apply the method to a realistic optimized scenario with $m = 7000$ iterations and a $256$-band strategy matrix with unit-norm columns optimized for mean squared error,\footnote{\texttt{toeplitz.optimize\_banded\_toeplitz(n=7000, bands=256)} in \texttt{jax\_privacy}.} balls-in-bins sampling over a single epoch ($T = 7000$), and a target of $(10, 10^{-9})$-DP. The noise multiplier is swept from $0.30$ to $0.41$; \cref{fig:banded256} shows the resulting sample counts in the format of \cref{fig:mixtures}; the operating point is $\sigma = 0.40$, where $\hat{\delta} = 1.2 \cdot 10^{-9}$.

The gain at the operating point is $1/B \approx 190\times$: smaller than deeper in the tail (at $\sigma = 0.42$ it is $450\times$, at $\sigma = 0.45$ about $1700\times$), because at $\sigma = 0.40$ a single mode's $\delta$ is only about $10^{-3}$ and amplification supplies the remaining six orders of magnitude, which $B$ does not see. Even so, importance sampling reduces the number of samples to certify from $5.6 \cdot 10^{10}$ to $2.9 \cdot 10^8$. As in \cref{tab:variance}, the estimator itself is far more precise than the certificate: estimating $\hat{\delta}$ to $10\%$ relative standard error took $1.8 \cdot 10^5$ samples at this point, and the empirical variance of the weighted samples is $2600$ times smaller than $B \hat{\delta}$. The reverse direction was again zero on every sample, and only the forward direction needed to be estimated to precision.

\paragraph{Future work.} The gap in \cref{tab:variance} is nonetheless real headroom, because the certification is one-sided. For a nonnegative estimand, a classical lower-tail bound \citep{maurer2003bound} controls the probability of under-estimating the mean in terms of the second moment $\mathbb{E}[Y^2]$ alone, with no dependence on the support; the trivial bound $\mathbb{E}[Y^2] \le B \delta$ recovers the rate of \cref{sec:samplecomplexity}, while \cref{tab:variance} indicates that the true second moment is $5.8$ to $1.7 \cdot 10^4$ times smaller. A proven bound on $\mathbb{E}_R[Y^2]$ for the proposal of \cref{sec:importancesampling}---an average over $\bfy$, rather than the pointwise supremum of \cref{lem:unified}---would therefore reduce the number of samples to certify by a corresponding factor. We leave this to future work.

\bibliographystyle{plainnat}
\bibliography{references}

\end{document}